\documentclass[onecolumn,nobibnotes,nofootinbib,superscriptaddress]{revtex4-2}
\usepackage{amsmath,amssymb,bm}
\usepackage[a4paper,bindingoffset=0.2in,left=0.8in,right=0.8in,top=1in,bottom=1in,footskip=.25in]{geometry}
\usepackage{graphicx,subfigure,epsfig}
\usepackage{bigints}
\usepackage{color}
\usepackage[breaklinks,colorlinks,urlcolor=blue,citecolor=blue,linkcolor=blue]{hyperref}
\usepackage{mciteplus}
\definecolor{lcolor}{rgb}{0.5,0,0}
\definecolor{citcolor}{rgb}{0,0.3,0.0}
\usepackage[capitalise]{cleveref}
\usepackage{todonotes}

\newcommand{\be}{\begin{equation}}
\newcommand{\ee}{\end{equation}}
\newcommand{\beq}{\begin{eqnarray}}
\newcommand{\eeq}{\end{eqnarray}}
\newcommand{\benn}{\begin{displaymath}}
\newcommand{\eenn}{\end{displaymath}}
\newcommand{\beann}{\begin{eqnarray*}}
\newcommand{\eeann}{\end{eqnarray*}}

\newcommand{\nn}{\nonumber\\}

\begin{document}

\title{Imaging constituent quark shape of proton with exclusive vector meson production at HERA}
\author{Wenchang Xiang}
\email{wxiangphy@gmail.com}
\affiliation{Physics Division, Guangzhou Maritime University, Guangzhou 510725, China}
\author{Yanbing Cai}
\email{myparticle@163.com}
\affiliation{Guizhou Key Laboratory in Physics and Related Areas, Guizhou University of Finance and Economics, Guiyang 550025, China}
\author{Daicui Zhou}
\email{dczhou@mail.ccnu.edu.cn}
\affiliation{Key Laboratory of Quark and Lepton Physics (MOE), and Institute of Particle Physics, Central China Normal University, Wuhan 430079, China}


\begin{abstract}
We show within proton hot spot picture that the exclusive vector meson production in electron-proton deeply inelastic scattering is sensitive to the individual width of the constituent quarks of the proton. For comparison, we calculate the exclusive $J/\Psi$ production cross-sections in three cases, $\mathrm{B_u} \geq \mathrm{B_d}$, $\mathrm{B_u} < \mathrm{B_d}$ and $\mathrm{B_u}\neq\mathrm{B^{\prime}_u}\neq\mathrm{B_d}$, where the $\mathrm{B_u}$, $\mathrm{B^{\prime}_u}$ and $\mathrm{B_d}$ denote the widths of two up quarks and a down quark. We find that only results calculated with $\mathrm{B_u} \geq \mathrm{B_d}$ can give a reasonable description of the exclusive $J/\Psi$ production cross-section data at HERA. To test that our results are independent of the details of the model, we retain the average width of the three constituent quarks unchanged and compute the exclusive $J/\Psi$ production cross-sections with contribution weight by setting different proportional coefficients ($\mathrm{W_u}$ and $\mathrm{W_d}$) for the up and down quarks, respectively. It shows that the results calculated with $\mathrm{W_u}\geq\mathrm{W_d}$ can well reproduce the exclusive $J/\Psi$ production data at HERA, while the opposite case cannot describe the HERA data. These interesting findings seem to indicate that the up quark has more gluons around it than the down quark at high energy although the spatial distribution of gluons fluctuates event-by-event . To ensure the relevant results independent of the species of the vector meson, we also calculate the $\rho$ production cross-sections with the same group of parameters used in the exclusive $J/\Psi$ production and compare the predictions with the HERA data. It shows that all the results computed in the exclusive $\rho$ productions are consistent with the findings obtained in the exclusive $J/\Psi$ productions.
\end{abstract}

\maketitle

\section{Introduction}
\label{intro}
Investigating the structure of the proton is a fundamental interest in high energy physics, which includes its radius, geometric shape and event-by-event fluctuations. It is well known that the deeply inelastic electron-proton ($e+p$) scattering is one of the most effective way to access the proton structure, which can provide the cleanest data to constrain the phenomenological models. For instance, the exclusive vector meson production cross-section data measured in H1 and ZEUS experiments at Hadron-Electron Ring Accelerator (HERA) have been successfully used to extract the gluonic root mean square radius of the proton\cite{Caldwell:2010zza} and find the hot spot picture of the proton\cite{Mantysaari:2016ykx,Mantysaari:2016jaz}.

There have been numerous studies to indicate that the Color Glass Condensate (CGC) effective field theory can provide a natural and easy way to describe the deeply inelastic scattering (DIS) process\cite{Golec-Biernat:1999qor,Iancu:2003ge,Watt:2007nr,Kozlov:2007wm,Albacete:2009fh,Rezaeian:2013tka,
Contreras:2016joy,Cepila:2018faq,Ducloue:2019jmy,Hanninen:2022gje}. In the CGC framework, the $\gamma^*+p$ DIS is described by the dipole picture\cite{Mueller:1993rr,Mueller:1994jq}, where the virtual photon fluctuates into a quark-antiquark dipole pair, then the dipole scatters off the proton target. In fact, the CGC framework has been widely used in studying proton structure functions at HERA energies, as well as a variety of observables at Relativistic Heavy Ion Collider (RHIC) and Large Hadron Collider (LHC) energies, for instance, the rapidity and transverse momentum distribution of the inclusive single and double hadron productions in both proton-proton and proton-nucleus collisions\cite{Kharzeev:2001gp,Dumitru:2005gt,Albacete:2010bs,Levin:2010zy}. In addition to the inclusive processes, the CGC framework has also been successfully extended to describe the exclusive diffractive processes\cite{Kowalski:2006hc,Marquet:2007qa,Goncalves:2009za,Armesto:2014sma,Mantysaari:2016ykx,Cai:2020exu,Demirci:2022wuy,
Mantysaari:2023qsq}.

The exclusive diffractive vector meson production is a powerful tool to probe the geometrical structure of the proton at small Bjorken-$x$, since the net color charge exchange is not allowed in the exclusive diffractive processes which require at least two gluons exchanged during the interaction, leading to the cross-section proportional to the squared parton distributions\cite{Ryskin:1992ui}. Furthermore, in vector meson production process there is a rapidity gap (a region in rapidity with no produced particles) between a produced vector meson and the proton, which provides a clean approach to identify diffractive events in experiments. Moreover, the exclusive diffractive process gives a possibility to determine the squared momentum change ($t$) of the proton. It is known that the $t$ is a Fourier conjugate of the impact parameter profile of the proton. Consequently, studying the exclusive diffractive process can provide access to the spacial structure of the proton.

In recent years, a lot of efforts were put into studying proton shape based on the exclusive diffractive process\cite{Schlichting:2014ipa,Mantysaari:2016ykx,Bendova:2022xhw}. Several hot spot models were proposed to investigate the event-by-event fluctuation of the spatial distribution of gluons within the proton\cite{Mantysaari:2016ykx,Mantysaari:2016jaz,Albacete:2016pmp,Cepila:2016uku,Traini:2018hxd,Kumar:2021zbn,Demirci:2021kya,Demirci:2022wuy}. All these models are inspired by the constituent quark (uud, here u and d represent up and down quarks) picture of the proton, thus they share basic physics ingredients. They assume that the large-$x$ constituent quarks are the sources of small-$x$ gluon, so the emitted gluons distribute around the valence quark to form ``gluon cloud'' also called hot spot. An early form of the hot spot model can trace back to Schlichting and Schenke's work in Ref.\cite{Schlichting:2014ipa} in which they assumed that the ``gluon cloud'' is around the proton's constituent quarks. A continuing work was done in Refs.\cite{Mantysaari:2016ykx,Mantysaari:2016jaz} where a formal hot spot model is formed, they found the incoherent exclusive diffractive $J/\Psi$ data at HERA can be well reproduced once the event-by-event geometric fluctuations of proton are included. If there is no hot spot contribution, the theoretical predictions of the incoherent cross-sections are several magnitude smaller than the experimental measurements. An analytical model for the hot spot structure of the proton was built in Refs.\cite{Demirci:2021kya,Demirci:2022wuy} where they took into account the non-relativistic and dilute limits and kept the model analytically tractable. Stimulated by the important impact of the target fluctuations, the projectile fluctuations were studied in Ref.\cite{Blaizot:2022bgd}, they found that the dipole size fluctuation plays an essential role in describing the exclusive $J/\Psi$ production data at very small $|t|$. They concluded that the dipole size fluctuation can be interpreted as fluctuation of the saturation scale of the proton. The authors in Ref.\cite{Cepila:2016uku} found that the number of the hot spot grows with decreasing $x$, where $x$ is the momentum fraction (or Bjorken $x$). A novel work about the substructure of the hot spot and the energy dependence of the proton's transverse size and the hot spot's transverse size was performed in Refs.\cite{Kumar:2021zbn,Kumar:2022aly}, they found that the substructure of the hot spot plays a significant role in the description of the incoherent cross-section of the exclusive $J/\Psi$ production at high $|t|$. Moreover, they also found that the transverse widths of the proton and hot spot increase as $x$ decreases.

All the aforementioned studies about the phenomenological applications of the hot spot model in the exclusive vector meson production assume that the three constituent quarks of proton have the same width of gluon distribution ($\mathrm{B_u}=\mathrm{B_d}=\mathrm{\bar{B}_{cq}}$), which means they used an average width to calculate the exclusive vector meson productions instead of individual width of the constituent quarks, note that the subscript ``$\mathrm{cq}$'' denotes constituent quark. We would like note point out that although authors in Ref.\cite{Kumar:2022aly} found that the width of the hot spot increases as $x$ decrease, in which the width of the hot spot is still an average width (not the width of an individual hot spot). In fact, for a physical event the hot spots can have different widths. It is very interesting to see what happens if the width of the gluon distribution around the u quark is different from the d quark.

Our aim in this paper is to study the possible width of gluon distribution around the constituent u and d quarks in proton at high energies using the $e+p\rightarrow e^{\prime}+p^{\prime}+V$ exclusive diffractive vector meson ($V$) productions. To this end, we use the hot spot model to calculate the exclusive vector meson productions with different widths for individual constituent quarks of the proton, but retaining the average width of the three constituent quarks unchanged. The average width is  extracted from the exclusive $J/\Psi$ production data in Ref.\cite{Mantysaari:2016jaz}. There are three classifications in terms of our scenario\footnote{In this paper, we assume that the two up quarks are not distinguishable in (1) and (2), but they are distinguishable in (3).}, (1) two up quarks having same width ($\mathrm{B_u}$) while the down quark having different width ($\mathrm{B_d}$) with $\mathrm{B_u} \geq \mathrm{B_d}$, (2) two up quarks having same width while the down quark having different width with $\mathrm{B_u} < \mathrm{B_d}$, (3) all three constituent quarks having different width ($\mathrm{B_u}\neq \mathrm{B^{\prime}_u}\neq \mathrm{B_d}$). Our results show that only case (1) can reproduce the incoherent exclusive vector meson production data measured at HERA, the predictions from cases (2) and (3) have a visible deviation from experimental data in the incoherent process. This interesting finding seems to indicate that the u quark has more gluons around it than the down quark at high energies. To ensure this outcome, we use another way to test it. We keep the individual width of the up and down quarks unchanged, but changing their contribution weight ($\mathrm{W_u}$ and $\mathrm{W_d}$) to the proton density profile. We would like to note that the larger contribution weight means more gluons surround the constituent quark. We compute the production cross-section of the exclusive vector meson again. We find that in the case of $\mathrm{W_u}\geq \mathrm{W_d}$ our calculations are in agreement with the exclusive vector meson production data at HERA, while the opposite case with $\mathrm{W_u} < \mathrm{W_d}$ cannot reproduce the data. This result also implies that the up quark with more gluons than the down quark is favorable by the exclusive vector meson production data at HERA. In addition to the exclusive $J/\Psi$ productions, the exclusive $\rho$ productions are used to verify the finding mentioned above. All the outcomes from the exclusive $\rho$ productions are consistent with the ones obtained in the exclusive $J/\Psi$ productions.


%
\begin{table}[htbp]
  \begin{center}
  \begin{tabular}{cc|cccccc}
  \hline
  &Meson &\quad $M_V/\text{GeV}$ &\quad $m_f/\text{GeV}$ &\quad $\mathcal{N}_{T}$ &\quad $\mathcal{N}_{L}$ &\quad $\mathcal{R}_{T}/\text{GeV}^{-2}$ &\quad $\mathcal{R}_{L}/\text{GeV}^{-2}$ \\
  \hline
  &  $J/\psi$  &\quad 3.097    &\quad 1.4    &\quad 0.578    &\quad  0.575    &\quad 2.3    &\quad 2.3 \\
  \hline
  &  $\rho$ &\quad 0.776   &\quad 0.14    &\quad 0.911    &\quad  0.853    &\quad  12.9    &\quad 12.9 \\
  \hline
    \end{tabular}%
  \caption{Parameters of the boosted Gaussian formalism for $J/\psi$ and $\rho$\cite{Kowalski:2006hc}.}
  \label{table:1}
  \end{center}
\end{table}%

\section{Theoretical framework of exclusive diffractive vector meson production}
To introduce notations and set up framework, we review the general aspects of the calculations of exclusive diffractive vector meson production in deeply inelastic scattering in this section.

\subsection{Exclusive vector meson production in dipole picture}
\label{dippic}
\begin{figure}[t!]
\setlength{\unitlength}{1.5cm}
\begin{center}
\epsfig{file=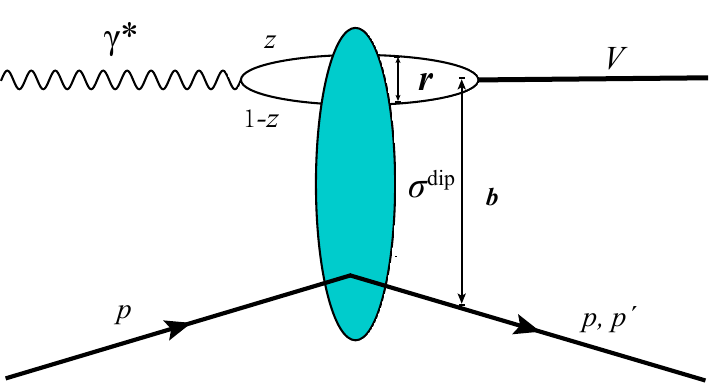, width=7cm,height=3.8cm}
\end{center}
\caption{The exclusive vector meson production in dipole picture.}
\label{vmp}
\end{figure}

One of natural and easy approaches to describe the exclusive vector meson production in the $\gamma^*p$ DIS is the dipole picture together with the CGC formalism. In dipole picture, the projectile virtual photon fluctuates into a quark-antiquark ($q\bar{q}$) dipole pair which then interacts with the color field in the target proton. After the interaction the $q\bar{q}$ dipole forms a vector meson in the final state of scattering, as shown in Fig.\ref{vmp}. The scattering amplitude for the exclusive vector meson production in the $\gamma^*p$ DIS can be expressed as\cite{Kowalski:2006hc}
\be
\mathcal{A}^{\gamma^*p\rightarrow Vp}_{T,L}(x, Q^2, \bm{\Delta}) = i \int d^2\bm{r}\int d^2\bm{b}\int\frac{dz}{4\pi}\big(\Psi_{\gamma}^*\Psi_V\big)_{T,L}(Q^2,\bm{r},z)\exp\Big\{-i\big[\bm{b}-(1-z)\bm{r}\big]\cdot\bm{\Delta}\Big\}
\frac{d\sigma^{\mathrm{dip}}}{d^2\bm{b}}(\bm{b},\bm{r},x),
\label{dipamp}
\ee
where the subscripts $T$ and $L$ denote the transverse and longitudinal polarizations of the virtual photon. The $\bm{r}$ and $\bm{b}$ represent the transverse size of $q\bar{q}$ dipole and the impact parameter of the dipole with respect to the proton target, respectively. The $z$ refers to the longitudinal momentum fraction of the quark. The $Q^2$ is the virtuality of the photon. Note that the transverse momentum transfer $\bm{\Delta}$ is the Fourier conjugate to $\bm{b}-(1-z)\bm{r}$. The $\sigma^{\mathrm{dip}}(\bm{b},\bm{r},x)$ is the dipole-proton cross-section which includes all the important information about the QCD dynamics. The details of the dipole cross-section will be discussed individually later in this section. The virtual photon wave function $\Psi_{\gamma}$ describes the splitting of $\gamma^*\rightarrow q\bar{q}$, which can be precisely computed by QED. The vector meson wave function $\Psi_V$ accounts for the formation of a vector meson from the scattered $q\bar{q}$ dipole, which is non-perturbative and needs to be modeled. The overlap function between the virtual photon and the vector meson has transverse and longitudinal components and can be written as\cite{Kowalski:2006hc}
\be
\label{eq:overlapT}
  (\Psi_\gamma^*\Psi_V)_{T} = \hat{e}_f e \frac{N_c}{\pi z(1-z)}\Big\{m_f^2 K_0(\epsilon r)\phi_T(r,z)
- [z^2+(1-z)^2]\epsilon K_1(\epsilon r) \partial_r \phi_T(r,z)\Big\},
\ee
\be
\label{eq:overlapL}
  (\Psi_\gamma^*\Psi_V)_{L} = \hat{e}_f e \frac{N_c}{\pi}2Qz(1-z)K_0(\epsilon r)
\Big[M_V\phi_L(r,z)+ \delta\frac{m_f^2 - \nabla_r^2}{M_Vz(1-z)}
    \phi_L(r,z)\Big],
\ee
where the effective charge $\hat{e}_f = 2/3, \mathrm{or}~1/\sqrt{2}$, for $J/\Psi$, or $\rho$ mesons respectively. The $K_0$ and $K_1$ in Eqs.(\ref{eq:overlapT}) and (\ref{eq:overlapL}) are the modified Bessel functions of the second kind with $\epsilon^{2}=z(1-z)Q^{2}+m_f^2$. In Eqs.(\ref{eq:overlapT}) and (\ref{eq:overlapL}), $\phi(r,z)$ is the scalar function. There are several models for the scalar function in the literature, such as Boosted Gaussian\cite{Nemchik:1994fp,Nemchik:1996cw}, Gauss-LC\cite{Kowalski:2003hm}, and NRQCD\cite{Lappi:2020ufv} wave functions. In this work, we will use the Boosted Gaussian wave function from Ref.\cite{Kowalski:2006hc}, since it successfully describes the vector meson production data at HERA. In boosted Gaussian formalism, the scalar functions are given by
\be
\label{eq:BGT}
\phi_{T}(r,z) = \mathcal{N}_{T} z(1-z)\exp\Big(-\frac{m_f^2 \mathcal{R}_{T}^2}{8z(1-z)}
- \frac{2z(1-z)r^2}{\mathcal{R}_{T}^2} + \frac{m_f^2\mathcal{R}_{T}^2}{2}\Big),
\ee
\be
\label{eq:BGL}
\phi_{L}(r,z) = \mathcal{N}_{L} z(1-z)\exp\Big(-\frac{m_f^2 \mathcal{R}_{L}^2}{8z(1-z)}
- \frac{2z(1-z)r^2}{\mathcal{R}_{L}^2} + \frac{m_f^2\mathcal{R}_{L}^2}{2}\Big).
\ee
The values of the parameters $M_V$, $m_f$, $\mathcal{N}_{T,L}$, and $\mathcal{R}_{T,L}$ in the above equations are given in Table~\ref{table:1}.
We would like to point out that there is a minor difference between the Boosted Gaussian and other wave functions coming from the overall normalization. All the wave functions give almost the same $t$ dependence of the exclusive vector meson production cross-section.

\subsection{Good-Walker formalism of exclusive vector meson production}
An elegant formalism for the diffractive exclusive vector meson production is the Good-Walker picture\cite{Good:1960ba}, in terms of which the diffraction in high energy limit of QCD can be described according to eigenstates of the scattering matrix off the proton target. Generally, one can divide the diffractive processes into two type of events in terms of target proton dissociation or not. If the scattered proton keeps intact after the scattering, it refers to the coherent diffraction. While the scattered proton breaks up, it refers to the incoherent diffraction. For the coherent diffraction, the cross-section is obtained by the square of the average of the amplitude\cite{Kowalski:2006hc}
\be
\frac{d\sigma^{\gamma^*p\rightarrow Vp}}{dt} = \frac{(1+\beta^2)R_g^2}{16\pi}\Big|\big\langle \mathcal{A}^{\gamma^*p\rightarrow Vp}(x,Q^2,\bm{\Delta})\big\rangle\Big|^2,
\label{cohamp}
\ee
where $\mathcal{A}^{\gamma^*p\rightarrow Vp}(x,Q^2,\bm{\Delta})$ is the scattering amplitude given in Eq.(\ref{dipamp}). The $(1+\beta^2)$ in Eq.(\ref{cohamp}) is to account for the real part correction to the scattering amplitude, since it is missed when the diffractive scattering amplitude was derived. Here, the $\beta$ is the ratio of the real to imaginary part of the scattering amplitude
\be
\beta = \tan\Big(\frac{\pi\delta}{2}\Big)
\ee
with
\be
\delta = \frac{\partial\ln\big(\mathcal{A}_{T,L}^{\gamma^*p\rightarrow Vp}\big)}{\partial\ln(1/x)}.
\ee
The $R_g$ is the skewness correction which is to account for the imbalance contribution between the two exchanging gluons in the diffractive dipole-target scattering. In this work, we will use the prescription from Ref.\cite{Kowalski:2006hc} to include the skewness correction
\be
R_g =\frac{2^{2\delta+3}}{\sqrt{\pi}}\frac{\Gamma(\delta+5/2)}{\Gamma(\delta+4)}.
\ee

From Eq.(\ref{cohamp}), one can see that the coherent cross-section is sensitive to the average target configuration, since it is obtained from the average over scattering amplitude. Thus, the coherent cross-section can only provide the overall information about the structure of the proton. To resolve the internal structure of the proton, one needs to use the incoherent cross-section.
The incoherent cross-section is proportional to the variance of the target profile, which can be obtained as the variance\cite{Miettinen:1978jb,Caldwell:2010zza}
\be
\frac{d\sigma^{\gamma^*p\rightarrow Vp}}{dt} = \frac{(1+\beta^2)R_g^2}{16\pi}\Bigg(\Big\langle\big|\mathcal{A}^{\gamma^*p\rightarrow Vp}(x,Q^2,\bm{\Delta})\big|^2\Big\rangle-\Big|\big\langle \mathcal{A}^{\gamma^*p\rightarrow Vp}(x,Q^2,\bm{\Delta})\big\rangle\Big|^2\Bigg).
\ee
It renders that the incoherent cross-section extremely sensitive to the details of the structure fluctuations of the proton.

\subsection{Dipole cross-section}
The most important ingredient in the scattering amplitude for the exclusive vector meson production in Eq.(\ref{dipamp}) is the dipole-proton cross-section, since it includes all the QCD dynamics of the interaction between $q\bar{q}$ dipole and proton target. In terms of the optical theorem, one knows that the dipole cross-section can be obtained by the forward dipole scattering amplitude $N$ as following
\be
\frac{d\sigma^{\mathrm{dip}}}{d^2\bm{b}}(\bm{b}, \bm{r}, x) = 2N(\bm{b}, \bm{r}, x).
\ee
In the CGC framework, the rapidity (or small-$x$) evolution of the forward dipole scattering amplitude $N$ can be performed by the JIMWLK\footnote{The JIMWLK is the abbreviation of Jalilian-Marian, Iancu, McLerran, Weigert, Leonidov, Kovner.}\cite{Balitsky:1995ub,Jalilian-Marian:1997qno,Jalilian-Marian:1997jhx,Iancu:2000hn,Ferreiro:2001qy} or Balitsky-Kovchegov (BK)\cite{Balitsky:1995ub,Kovchegov:1999ua} evolution equations. In the past fifteen years, there was significant progress in the BK evolution equations beyond the leading order, the BK equation receives corrections from the quark loops (running coupling effect)\cite{Balitsky:2006wa}, gluon loops\cite{Balitsky:2007feb,Lappi:2015fma}, the tree gluon diagrams with quadratic and cubic nonlinearities, double logarithmic resummations\cite{Iancu:2015vea,Xiang:2019kre}, as well as transformation of the representation of rapidity\cite{Ducloue:2019ezk,Xiang:2021rcy}, which makes that the improved BK equation can provide an excellent description of the HERA data\cite{Ducloue:2019jmy,Beuf:2020dxl}.
However, we will not use the BK evolution equation in this work, since the impact parameter dependence of the dipole scattering amplitude $N$ is a key aspect to calculate the diffractive scattering amplitude in Eq.(\ref{dipamp}), the numerical studies of the JIMWLK and BK equations showed that their numerical solutions develop unphysical Coulomb tails once the impact parameter dependence is included\cite{Golec-Biernat:2003naj,Berger:2011ew}. Alternatively, the rapidity evolution of the forward dipole scattering amplitude $N$ can be modeled along with the $Q^2$ and impact parameter $\bm{b}$ dependence, as what have been done in the impact parameter dependent saturation model (IPsat)\cite{Kowalski:2003hm} and the impact parameter dependent Color Glass Condensate model (b-CGC)\cite{Watt:2007nr}. In this work, we choose to use the IPsat model to evolve the dipole scattering amplitude, since it has been successfully used to describe a variety of the DIS data at HERA, for instance, the proton structure functions $F_2$ and $F_L$ data\cite{Rezaeian:2013tka}.

In the IPsat model, the dipole cross-section is written as
\beq
\frac{d\sigma^{\mathrm{dip}}}{d^2\bm{b}}(\bm{b}, \bm{r}, x) &=& 2N(\bm{b}, \bm{r}, x)\nn
&=& 2\bigg[1-\exp\Big(-\frac{\pi^2\bm{r}^2}{2N_c}\alpha_s(\mu^2)xg(x,\mu^2)T_p(\bm{b})\Big)\bigg],
\label{ipsat}
\eeq
where the $T_p(\bm{b})$ is the proton profile function and is assumed to be Gaussian in this work
\be
T_{p}(\bm{b})=\frac{1}{2\pi \mathrm{B_p}}\exp\bigg(-\frac{\bm{b}^2}{2\mathrm{B_p}}\bigg),
\ee
with $\mathrm{B_p}$ to be as the proton width.
The $xg(x,\mu^2)$ in Eq.(\ref{ipsat}) is the gluon density which evolves up to $\mu$ with the DGLAP gluon evolution. Here, the $\mu$ is related to the transverse dipole size as
\be
\mu^2 = \frac{4}{\bm{r}^2} + \mu_0^2,
\ee
and the initial $xg(x,\mu^2)$ at $\mu_0^2$ is
\be
xg(x,\mu_0^2) = A_g x^{-\lambda_g}(1-x)^{5.6},
\ee
where the $\mu_0$, $A_g$, and $\lambda_g$ are model parameters which were determined by fitting to the HERA DIS data in Ref.\cite{Rezaeian:2012ji}.

\begin{figure}[t!]
\setlength{\unitlength}{1.5cm}
\begin{center}
\epsfig{file=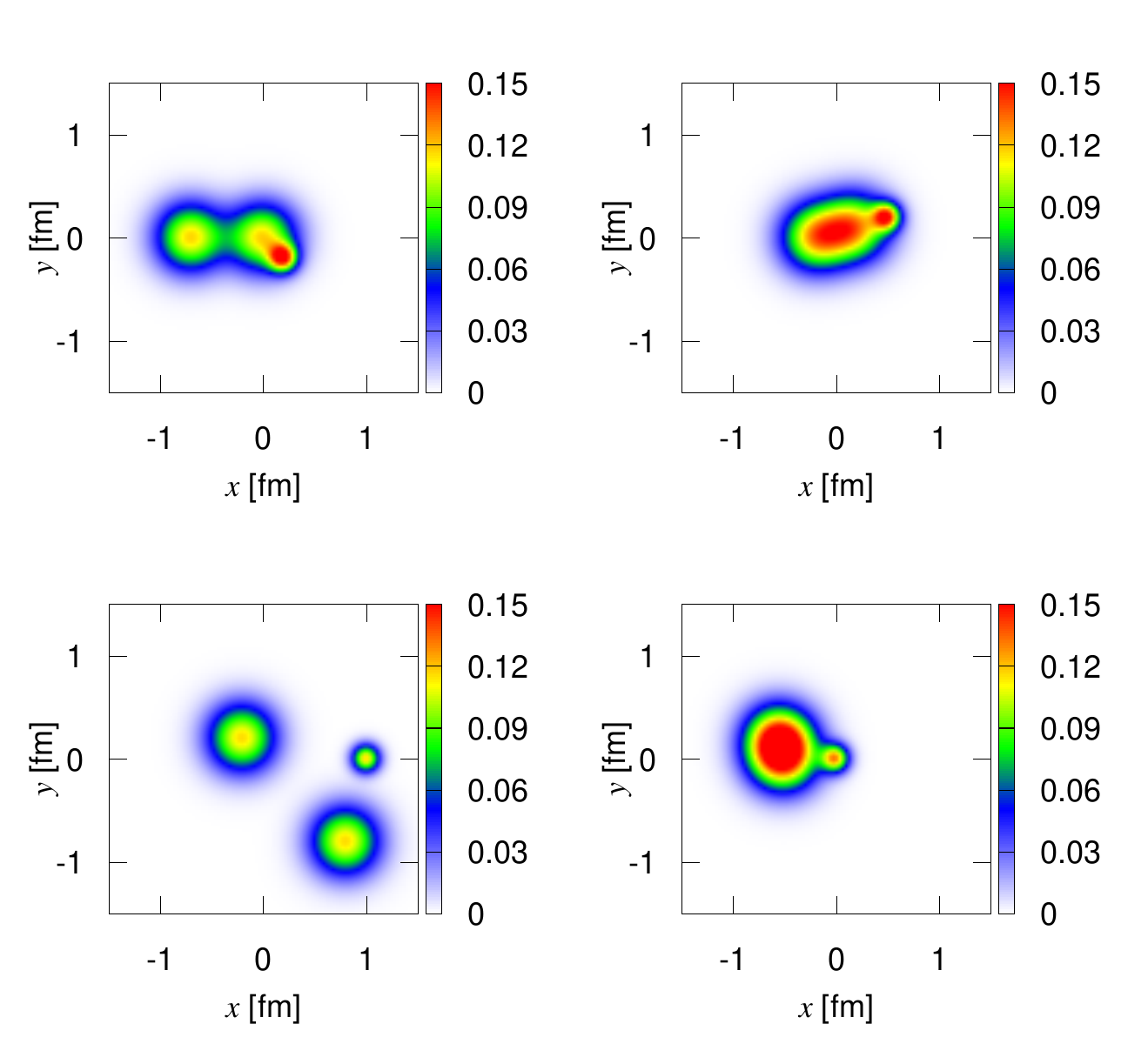, width=12cm,height=10.2cm}
\end{center}
\caption{Selected proton density profiles at $x\sim10^{-3}$ with $\mathrm{\bar{B}_{cq}}=1.0$ and $\mathrm{B_{qp}}=3.0$.}
\label{fig2}
\end{figure}
%

\section{Hot spot model}
\label{sec:hsm}
It is known that the geometric shape of a proton fluctuates event-by-event\cite{Schlichting:2014ipa,Mantysaari:2016ykx,Mantysaari:2016jaz}. A general approach to describe the fluctuating shape of the proton's gluon distribution is the hot spot model which was inspired by the constituent quark picture. In hot spot model\cite{Mantysaari:2016ykx,Mantysaari:2016jaz}, the large-$x$ valence quarks are thought to be as the sources of the small-$x$ gluons. The radiated gluons stay around their own parent valence quarks to form ``gluon cloud'' also called hot spot.

In terms of the hot spot model the transverse positions ($\bm{b_i}$) of the constituent quarks respect to the origin are sampled by a Gaussian distribution with width ${\mathrm{B_{qp}}}$, in which the subscript ``$\mathrm{qp}$'' means quark position. The density profile of each constituent quark is also assumed to have Gaussian distribution
\be
T_{\mathrm{cq}}(\bm{b}) = \frac{1}{2\pi \mathrm{B_{cq}}}\exp\bigg(-\frac{\bm{b}^2}{2{\mathrm{B_{cq}}}}\bigg),
\ee
with $\mathrm{B_{cq}}$ to be the width of the constituent quark, specifically $\mathrm{B_{cq}}$ can be $\mathrm{B_u}$ for up quark or $\mathrm{B_d}$ for down quark. Correspondingly, the proton density profile in Eq.(\ref{ipsat}) should be replaced by\cite{Mantysaari:2016ykx,Mantysaari:2016jaz}
\be
T_p(\bm{b}) = \frac{1}{N_{hs}}\sum_{i=1}^{N_{hs}}\mathrm{W_i}T_{\mathrm{cq}}\big(\bm{b}-\bm{b_i}\big),
\label{bdep}
\ee
where the $N_{hs}$ is the number of hot spots, in general one can set $N_{hs}=3$ in terms of the number of the constituent quarks in a proton. It has been shown that the change of $N_{hs}$ takes a negligible impact on the diffractive cross-section of the exclusive $J/\Psi$ production\cite{Mantysaari:2016jaz}. Although we notice that the authors in Refs.\cite{Cepila:2016uku,Kumar:2022aly} found $N_{hs}$ having an energy dependence, we shall use $N_{hs}=3$ in the numerical studies in next section, since this work focuses on measurements at HERA energies. $\mathrm{W_i}$ in Eq.(\ref{bdep}) is the contribution weight which has the same meaning as the $\mathrm{W_u}$ or $\mathrm{W_d}$ mentioned in Sec.\ref{intro}.

The new ingredient of this study is that we consider the difference of the gluon distribution between the up quark and down quark. So, the width $\mathrm{B_u}$ and $\mathrm{B_d}$ can take different values instead of the average value of the width of the three constituent quarks as done in Ref.\cite{Mantysaari:2016ykx,Mantysaari:2016jaz}. By this way, one can investigate the individual shape of each constituent quark of the proton at high energies. We call this new interpretation model as the refined hot spot model in this paper.

In order to see the density profile of each constituent quark, we present four selected proton profiles in Fig.\ref{fig2} which are drawn by using the parameters extracted from the exclusive $J/\Psi$ production data in next  section. At first glance, one can see the ``lumpy'' structure of the proton. When one see the details of the proton, one can find that the size of two hot spots are larger than the other one, especially from the left hand side panel of Fig.\ref{fig2}. Here, we just present the profiles which have two hot spots larger than other one\footnote{We assume that the two larger hot spots have same width and are induced from the emission of the two up quarks of the proton.}, since we will show in the next section that the structures as in Fig.\ref{fig2} are the ones which can describe the exclusive vector meson production data at HERA. The other configurations of the hot spots, for example, ($\mathrm{B_u}<\mathrm{B_d}$) or ($\mathrm{B_u}\neq \mathrm{B^{\prime}_{u}}\neq \mathrm{B_d}$), cannot describe the exclusive vector meson production data at HERA. These outcomes seem to indicate that the constituent up quark of the proton is preference to emit more gluons than the down quark of the proton at high energies.

In addition to the geometric fluctuations of the constituent quark mentioned above, there is another source of fluctuations due to the fluctuations in saturation scale, which was found to be significant in the description of the incoherent cross-section of the exclusive $J/\Psi$ productions at low $|t|$ region. We shall include the saturation scale fluctuations by following Ref.\cite{Mantysaari:2016jaz}, in which the saturation scale fluctuates in terms of
\be
P\big(\ln Q_s^2/\langle Q_s^2\rangle\big)=\frac{1}{\sqrt{2\pi}\sigma}\exp\Bigg[-\frac{\ln^2Q_s^2/\langle Q_s^2\rangle}{2\sigma^2}\Bigg].
\label{lognorm}
\ee
It is known that the expectation value of log-normal distribution, Eq.(\ref{lognorm}), is
\be
E\big[Q_s^2/\langle Q_s^2\rangle\big]=\exp\big[\sigma^2/2\big],
\label{E_qs}
\ee
which renders the average of $Q_s^2$ about $13\%$ (for $\sigma=0.5$) larger than the one in the case of no saturation scale fluctuations. Therefore, we need to normalize the distribution function by the expectation value, (\ref{E_qs}), in order to keep the desired expectation value unchanged.

\section{Numerical results}
We give the numerical results of the coherent and incoherent $J/\Psi$ and $\rho$ productions calculated from the refined hot spot model with the constituent up and down quarks taking individual width instead of an average width. The results are present in terms of the width of the individual constituent quarks, which are divided into
three cases: (1).$\mathrm{B_u}\geq \mathrm{B_d}$, (2).$\mathrm{B_u}<\mathrm{B_d}$, and (3).$\mathrm{B_u}\neq \mathrm{B^{\prime}_u}\neq \mathrm{B_d}$.

\subsection{$J/\Psi$ production}
It is known that the exclusive $J/\Psi$ production is a good probe to resolve the structure of the proton, since its mass is large enough to make the perturbative calculation of its photoproduction cross-section reliable. In this subsection, we will study the coherent and incoherent cross-sections of the exclusive $J/\Psi$  productions by our refined hot spot model. In order to explore the fine structure of the constituent quark of the proton at high energy, we vary the quark width parameters $\mathrm{B_u}$ and $\mathrm{B_d}$, but retaining the average width of the constituent quarks $\bar{\mathrm{B}}_{\mathrm{cq}}=(2\mathrm{B_u}+\mathrm{B_d})/3 = 1.0$ unchanged. We compare our numerical results with the measurements from H1 collaboration at HERA at $\sqrt{s}=75~\mathrm{GeV}$\cite{H1:2013okq} which is corresponding to $x\sim10^{-3}$ where our CGC framework is valid to apply. It has been found that the hot post model with only geometric shape fluctuations underestimates the exclusive vector meson productions in low $|t|$ region\cite{Mantysaari:2016ykx}. In order to reproduce the HERA data, we include the saturation scale fluctuations by using the log-normal distribution given in Sec.\ref{sec:hsm}.

\begin{figure}[t!]
\setlength{\unitlength}{1.5cm}
\begin{center}
\epsfig{file=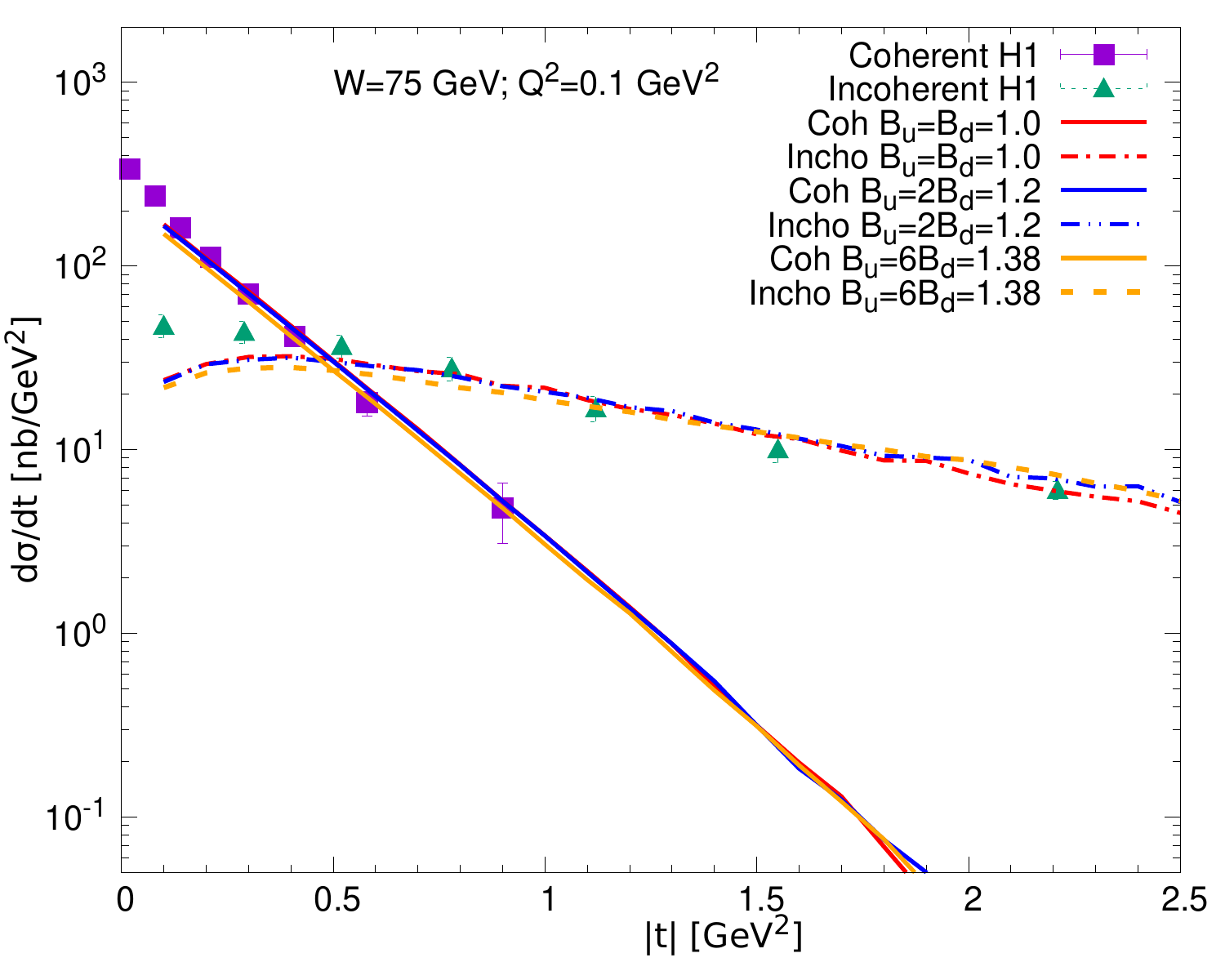, width=8.1cm,height=7.2cm}
\epsfig{file=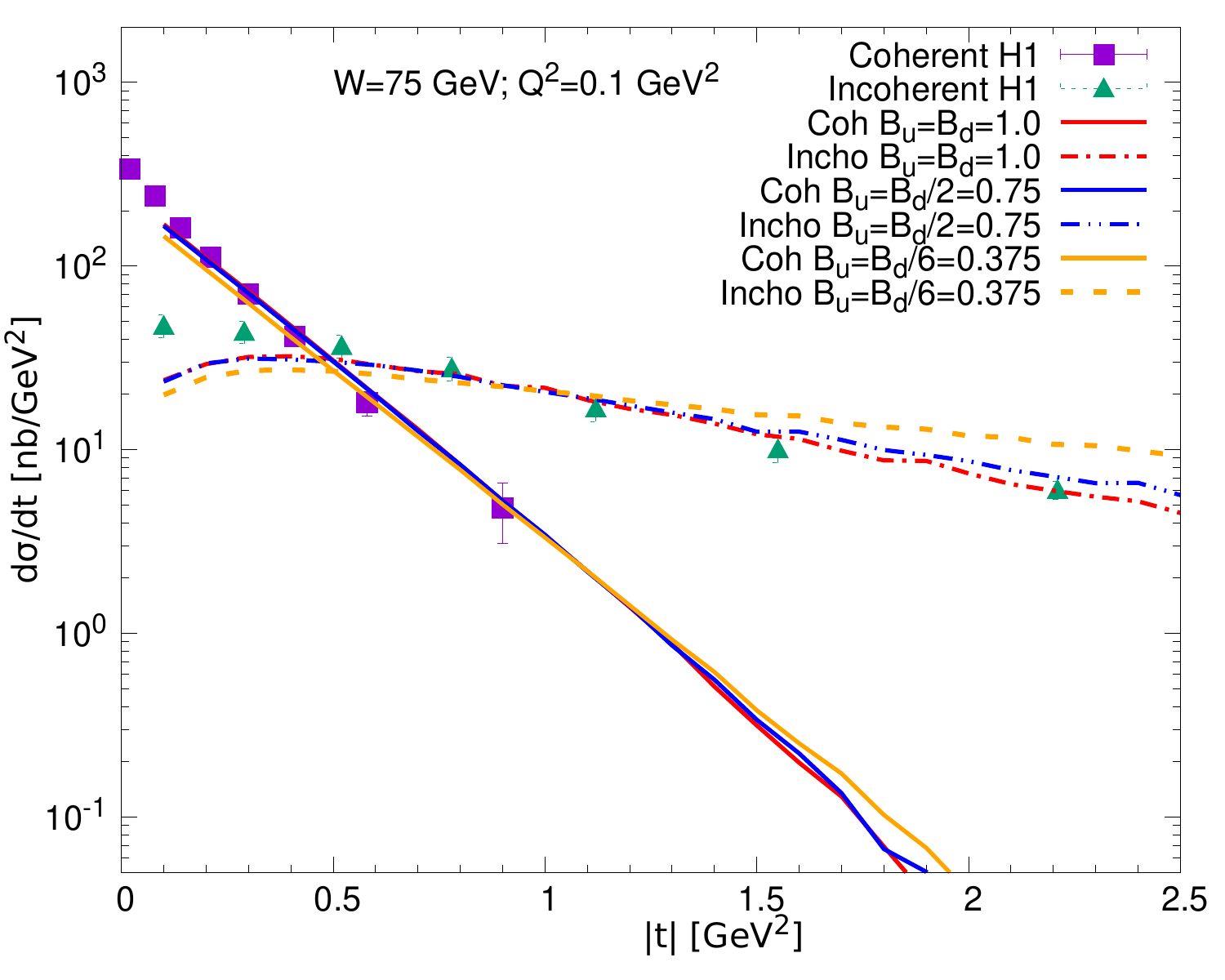, width=8.1cm,height=7.2cm}
\end{center}
\caption{The cross-sections of the coherent (solid curves) and incoherent (dashed curves) $J/\Psi$ production at $\sqrt{s}=75~\mathrm{GeV}$ as compared to the data from H1 collaboration\cite{H1:2013okq}. The left hand side panel shows the results calculated with $\mathrm{B_u} \geq \mathrm{B_d}$, while the results on right hand side panel computed with $\mathrm{B_u} < \mathrm{B_d}$.}
\label{fig3}
\end{figure}

The coherent and incoherent cross-sections of the exclusive $J/\Psi$ production as a function of momentum transfer $|t|$ at $\sqrt{s}=75~\mathrm{GeV}$ and $Q^2=0.1~\mathrm{GeV}$ are shown in Fig.\ref{fig3}. Note that the solid curves denote the numerical results of the coherent cross-sections, while the dashed curves represent the numerical results of the incoherent cross-sections (similarly hereinafter in following figures). The red curves are calculated by using the original hot spot model with $\mathrm{B_{qp}}=3.0~\mathrm{GeV}^{-2}$ and $\bar{\mathrm{B}}_{\mathrm{cq}}=1.0~\mathrm{GeV}^{-2}$\cite{Mantysaari:2016jaz}. We would like to note that in order to reproduce the data the width parameters $\mathrm{B_{qp}}$ and $\bar{\mathrm{B}}_{\mathrm{cq}}$ have tiny difference from the ones in the original hot spot model, which is not the focus of this study. We believe that the tiny difference should not affect the outcomes of this work. We use the red curves to be as the reference to evaluate the compatibility of the parameters of our refined hot spot model, since the red curves give rather good descriptions of the HERA data\cite{Mantysaari:2016jaz}. Figure.\ref{fig3} presents the comparisons of the coherent and incoherent cross-sections calculated by our refined hot spot model using five sets of width parameters. For the coherent process, one can see that all the results (solid curves) are good in agreement with the HERA data, there is no impact resulting from the fine structure difference among the constituent quarks of the proton, since the coherent cross-section is obtained by doing average on the level of scattering amplitude. Thus, it can only probe the average structure of the proton (rather than the fine structure of the proton), which is the reason why all the results about the coherent cross-sections in the following figures in this section are also in agreement with the HERA data in spite of the different width of the constituent quark applied.

To resolve the internal structure of a proton, one has to study the exclusive vector meson production in incoherent process, since the incoherent cross-section is proportional to the variance of the proton profile. For the incoherent process in Fig.\ref{fig3}, one can see that all results computed with $\mathrm{B_u} \geq \mathrm{B_d}$ are compatible with the measured cross-sections (see the left hand side panel in Fig.\ref{fig3}), while the numerical results cannot reproduce the HERA data when we use $\mathrm{B_u} < \mathrm{B_d}$, especially at large $|t|$ (see the right hand side panel in Fig.\ref{fig3}). This outcome seems to indicate that the up quark has more gluons around it than down quark at high energy although the spatial distribution of gluons fluctuates event-by-event. Note that in Fig.\ref{fig3} we only show the calculations with $\mathrm{B_u}=2\mathrm{B_d}$ and $\mathrm{B_u}=6\mathrm{B_d}$ in order to make the figure clearly visible. In fact, we also checked other cases, for example $\mathrm{B_u}=4\mathrm{B_d}$ and $\mathrm{B_u}=8\mathrm{B_d}$, all the results computed with $\mathrm{B_u}>\mathrm{B_d}$ are in agreement with the exclusive $J/\Psi$ production HERA data, since the gluon distribution shape of the constituent quark fluctuates event-by-event at high energies, which renders to the possibility that the width of the up quark can have any times of size larger than the width of down quark theoretically, in event-by-event cases.

\begin{figure}[t!]
\setlength{\unitlength}{1.5cm}
\begin{center}
\epsfig{file=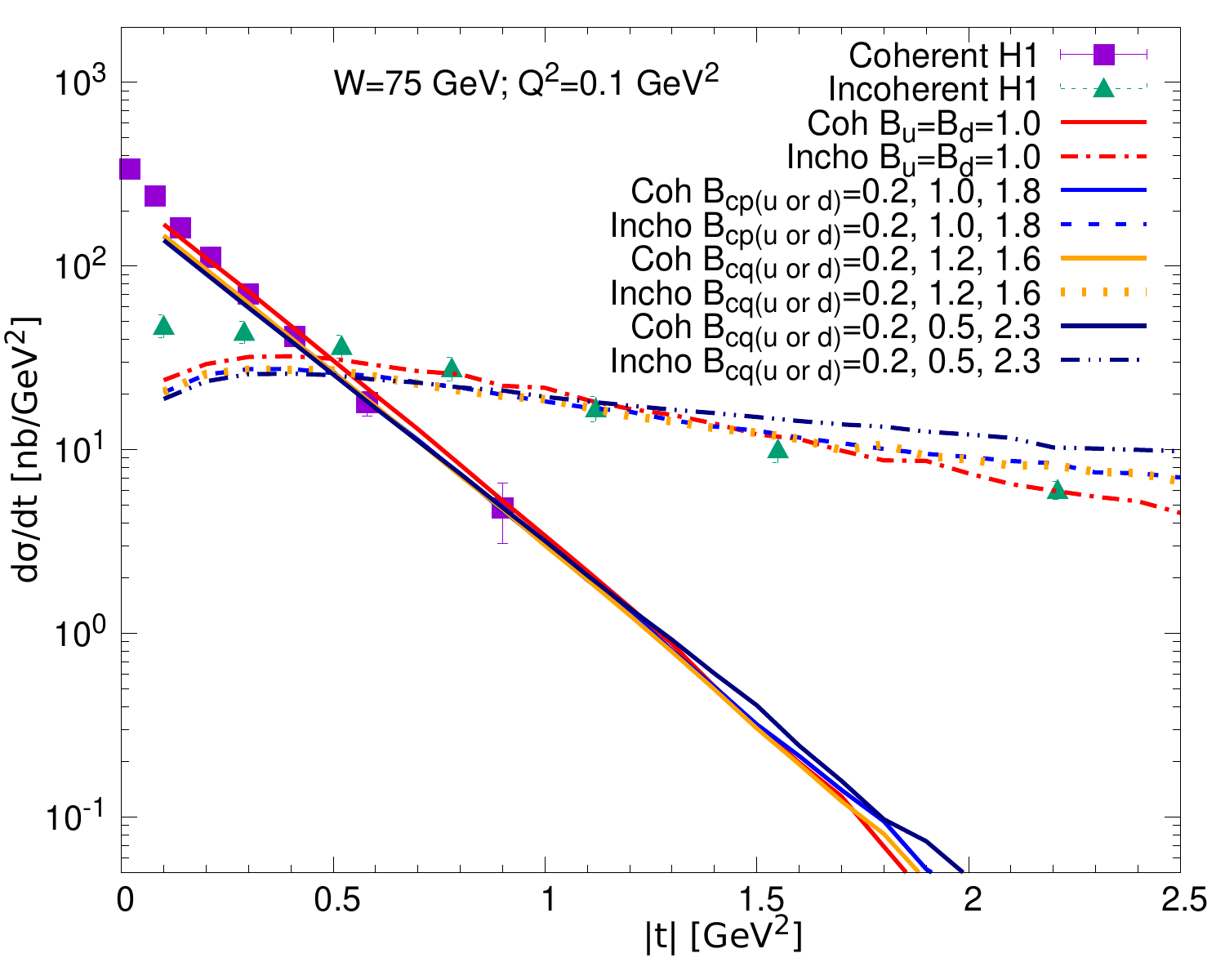, width=8.1cm,height=7.2cm}
\end{center}
\caption{The cross-sections of the coherent (solid curves) and incoherent (dashed curves) $J/\Psi$ production at $\sqrt{s}=75~\mathrm{GeV}$ as compared to the data from H1 collaboration\cite{H1:2013okq}. All the results are computed in the case of $\mathrm{B_u}\neq \mathrm{B^{\prime}_u}\neq \mathrm{B_d}$.}
\label{fig4}
\end{figure}
\begin{figure}[t!]
\setlength{\unitlength}{1.5cm}
\begin{center}
\epsfig{file=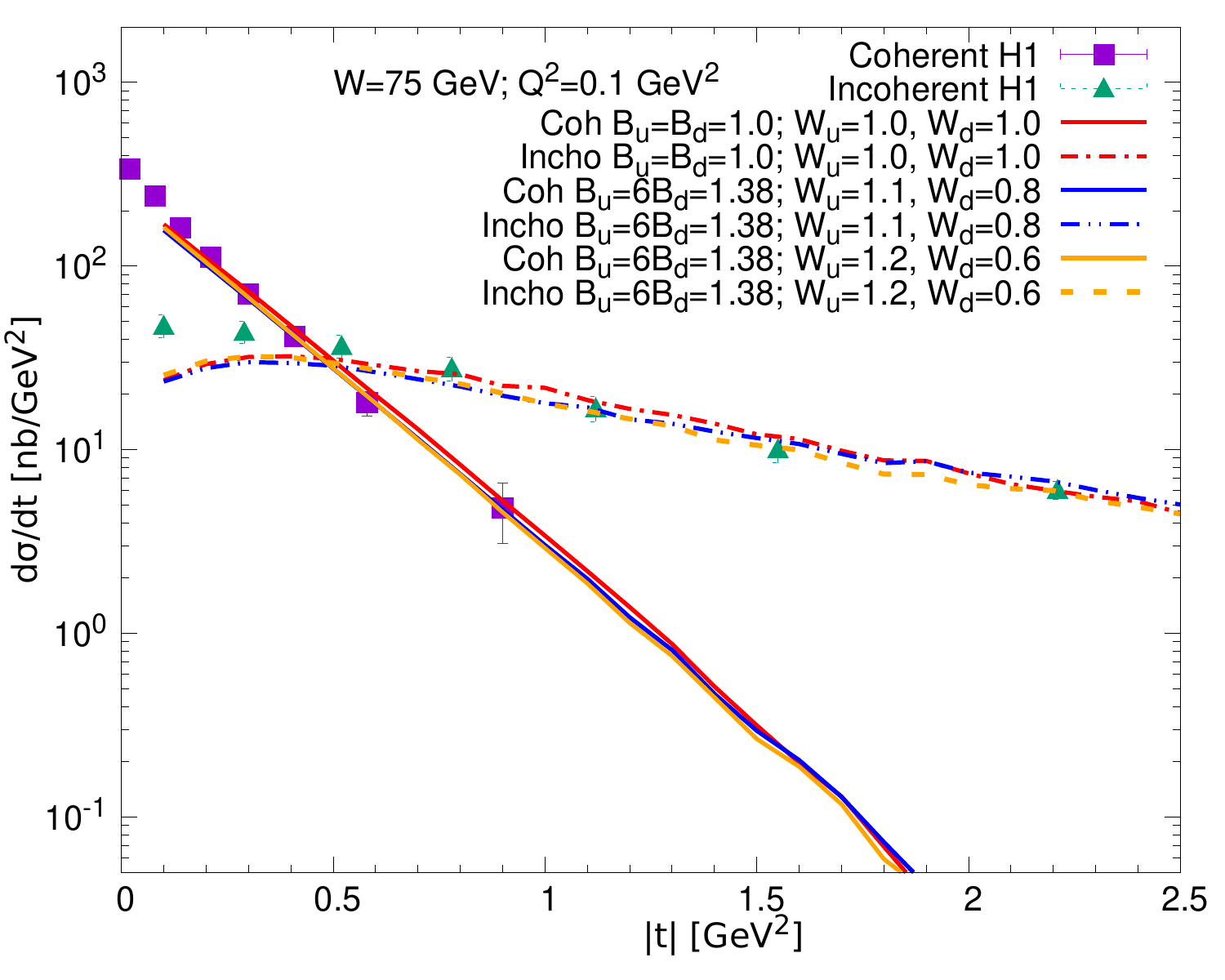, width=8.1cm,height=7.2cm}
\epsfig{file=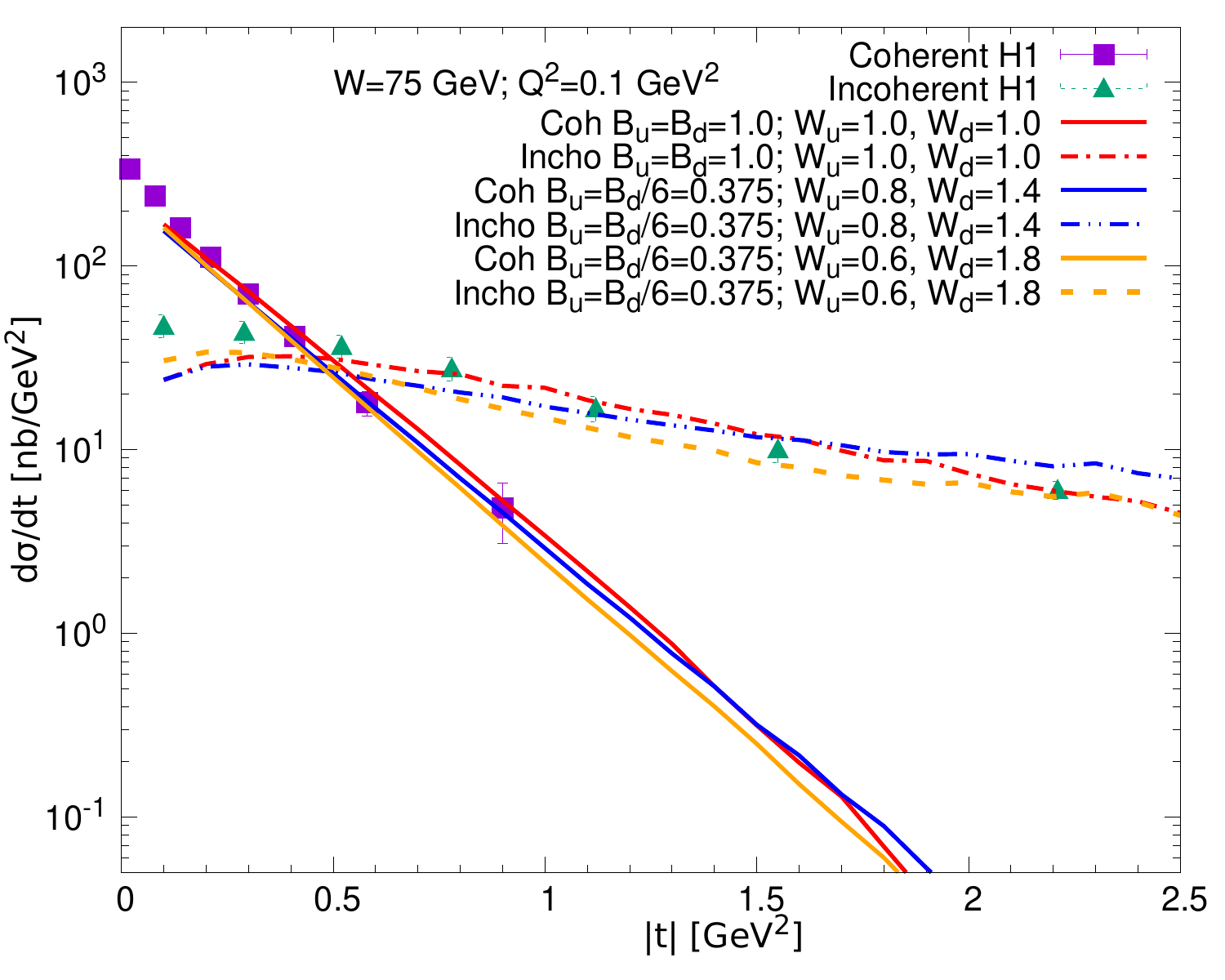, width=8.1cm,height=7.2cm}
\end{center}
\caption{The coherent (solid curves) and incoherent (dashed curves) cross-sections of $J/\Psi$ production at $\sqrt{s}=75~\mathrm{GeV}$ as a function of $|t|$ calculated by setting five sets of contribution weights ($\mathrm{W_u}$ and $\mathrm{W_d}$) compared to the data from H1 collaboration\cite{H1:2013okq}.The left hand side panel shows that the results computed with $\mathrm{B_u}\geq\mathrm{B_d}$ and $\mathrm{W_u}\geq\mathrm{W_d}$, while the right hand side panel demonstrates that results calculated with $\mathrm{B_u}<\mathrm{B_d}$ and $\mathrm{W_u}<\mathrm{W_d}$.}
\label{fig5}
\end{figure}

To ensure that the $\mathrm{B_u} > \mathrm{B_d}$ is the only one which is favorable by the HERA data, we perform calculations by using three different widths for the three constituent quarks of the proton. In order to get the consistent results as shown in Fig.\ref{fig3}, here we retain the average width of the constituent quarks unchanged, $\mathrm{\bar{B}}_\mathrm{{cq}}=1$. The relevant results are given in Fig.\ref{fig4} which shows the comparisons of the three selected calculations with the reference curves and HERA data. As expected, the data of the coherent cross-sections of the exclusive $J/\Psi$ productions are well reproduced by all the three cases. However, one can see that the incoherent cross-section data of the exclusive $J/\Psi$ productions cannot be well described by our refined hot spot model in the case of $\mathrm{B_u}\neq \mathrm{B^{\prime}_u}\neq \mathrm{B_d}$, which give us a hint that the gluon distributions of the constituent quarks with three different width could not exist in the proton at high energies even though the proton shape fluctuates event-by-event.

We also use another approach by using different contribution weight ($\mathrm{W_u}$ and $\mathrm{W_d}$) of the constituent quarks, to verify the findings obtained by using the individual widths of the constituent quarks. Note that the larger contribution weight means more gluons surround the constituent quark. In Fig.\ref{fig5}, we present the coherent and incoherent cross-sections of the exclusive $J/\Psi$ productions as a function of momentum transfer $|t|$, here the typical results with $\mathrm{B_u}=6\mathrm{B_d}$ and $\mathrm{B_u}=\mathrm{B_d}/6$ are selected. In our calculations, we retain the average contribution weight unchanged, $\mathrm{\bar{W}}_\mathrm{{cq}}=(2\mathrm{W_u}+\mathrm{W_d})/3=1.0$. From the left hand side panel of Fig.\ref{fig5}, one can see that the calculations with three sets contribution weights ($\mathrm{W_u}\geq\mathrm{W_d}$) can reproduce the data of the incoherent cross-section of the exclusive $J/\Psi$ productions as long as $\mathrm{B_u}>\mathrm{B_d}$, which seems to imply that the up quark has more gluons around it than down quark at high energy although the spatial distribution of gluons fluctuates event-by-event. The right hand side panel of Fig.\ref{fig5} shows the results calculated with $\mathrm{W_u}<\mathrm{W_d}$ and $\mathrm{B_u}<\mathrm{B_d}$, one can see that the results of the incoherent cross-section are not able to give a reasonable description of the HERA data almost in all $|t|$ regions. Finally, Fig.\ref{fig5} shows that the variance of the contribution weight does not affect the description of the coherent cross-section of the exclusive $J/\Psi$ production in $\gamma^*p$ DIS, since the coherent cross-section can only reflect the average shape of the proton.
To conclude, although Figs.(\ref{fig3}),(\ref{fig4}) and (\ref{fig5}) use two different approaches to evaluate the exclusive $J/\Psi$ productions, they show the same outcome ($\mathrm{B_u}\geq\mathrm{B_d}$), which indicates that our findings are independent of the model details.

\begin{figure}[t!]
\setlength{\unitlength}{1.5cm}
\begin{center}
\epsfig{file=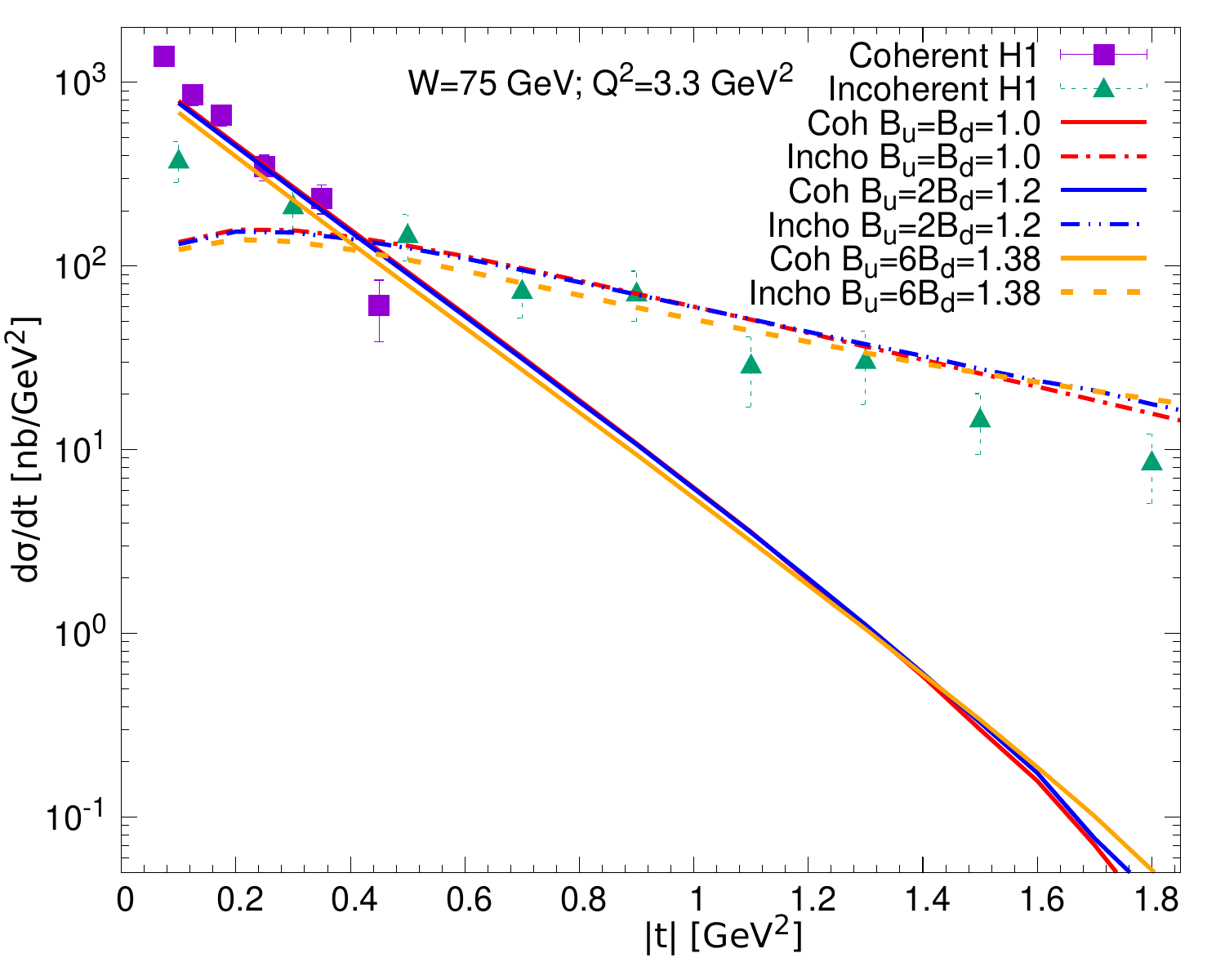, width=8.1cm,height=7.2cm}
\epsfig{file=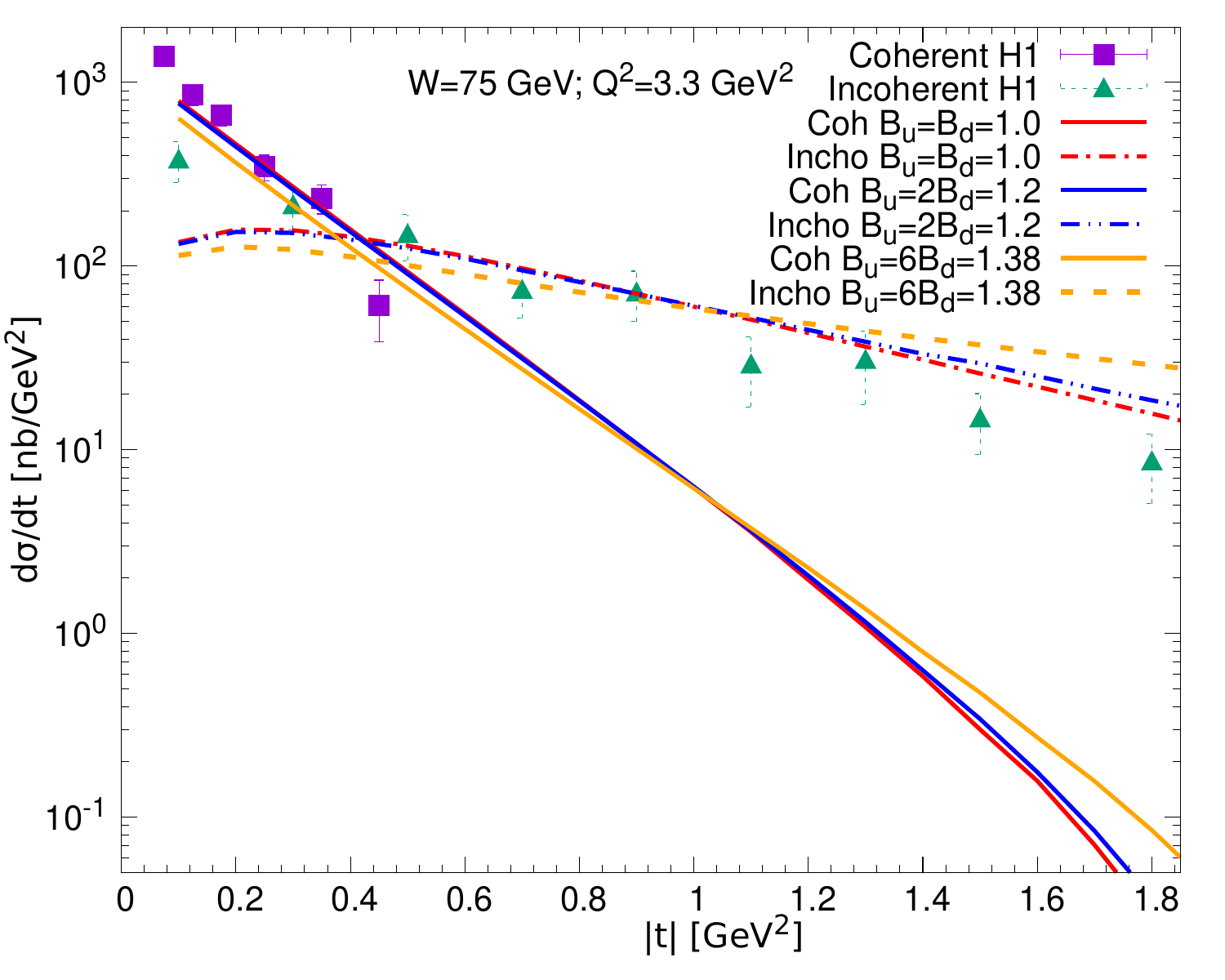, width=8.1cm,height=7.2cm}
\epsfig{file=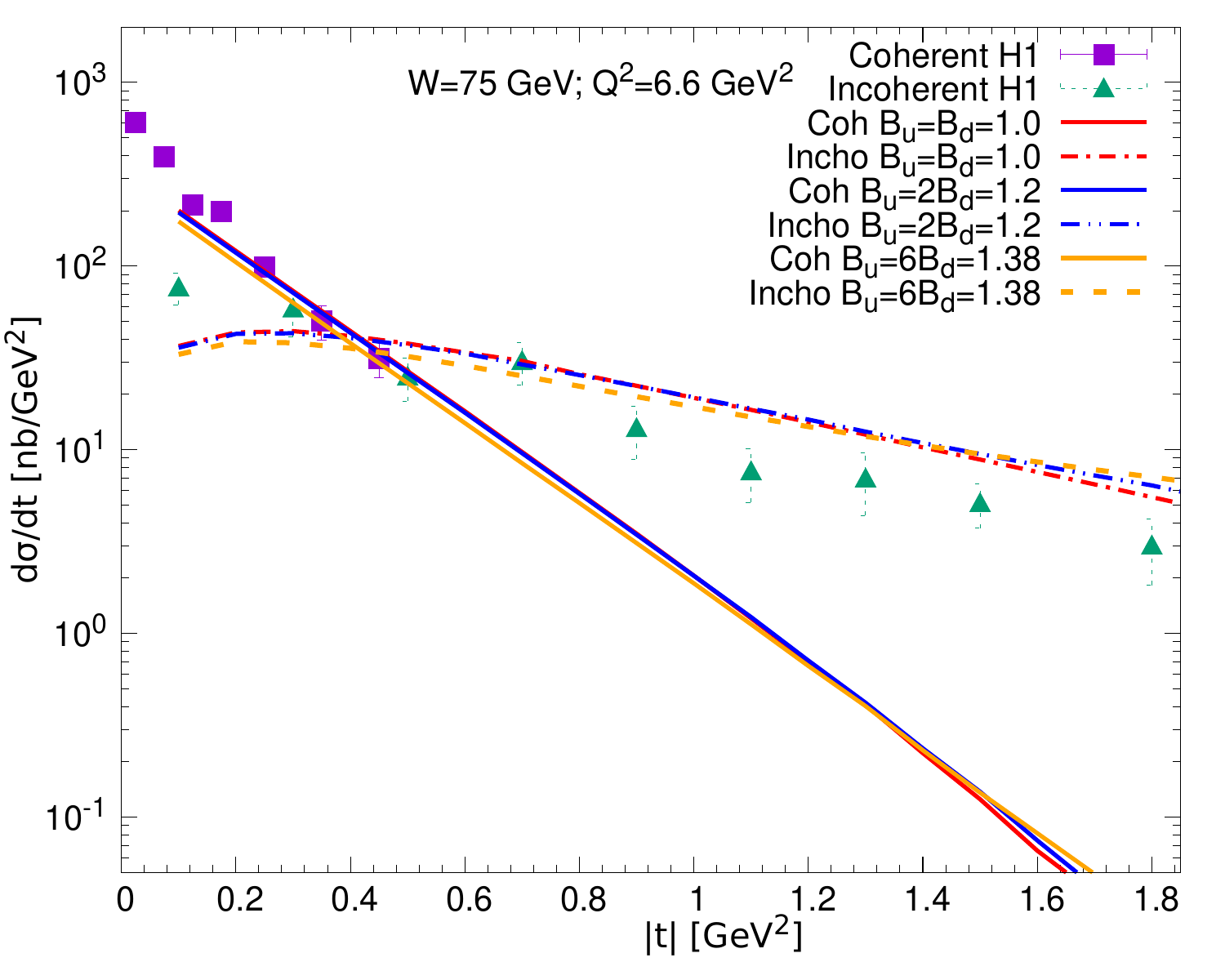, width=8.1cm,height=7.2cm}
\epsfig{file=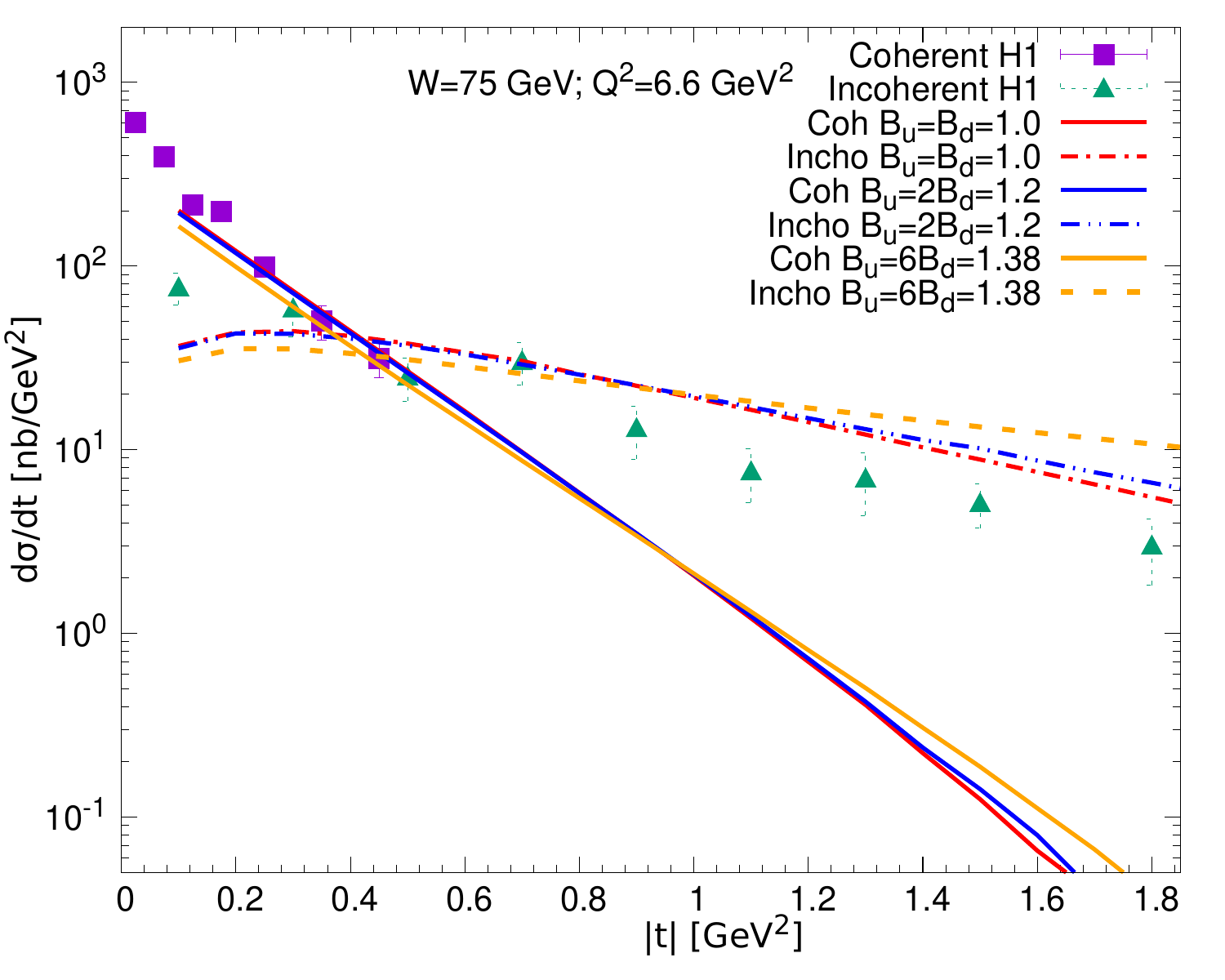, width=8.1cm,height=7.2cm}
\end{center}
\caption{The cross-sections of the coherent (solid curves) and incoherent (dashed curves) $\rho$ production at $\sqrt{s}=75~\mathrm{GeV}$ and $Q^2=3.3~\mathrm{GeV}^2$ and  $Q^2=6.6~\mathrm{GeV}^2$ as compared to the data from H1 collaboration\cite{H1:2009cml}. All the parameters used in the calculations are the same as Fig.\ref{fig3} except the vector meson changed from $J/\Psi$ to $\rho$.}
\label{fig6}
\end{figure}

\subsection{$\rho$ production}
Except $J/\Psi$, another alternative candidate to probe the internal structure of the proton is $\rho$ production in $\gamma^*p$ DIS. The exclusive $\rho$ productions have been measured at HERA\cite{H1:2009cml}, which provides access to explore the structure of the proton. However, we need to point out that the theoretical calculations of the exclusive $\rho$ production in the CGC framework has less reliability as compared to the exclusive $J/\Psi$ production, since the small mass of the $\rho$ makes its production cross-section receiving significant non-perturbative contributions at large dipole size. Although the dipole scattering amplitude (IPsat model), what we used in the calculations, also includes some non-perturbative physics by requiring the amplitude satisfying the unitarity limit, the IPsat model might be still out of its applicable region when the dipole size becomes large.  Therefore, we shall investigate the exclusive $\rho$ productions at large enough $Q^2$ in order to allow the perturbative treatment of its production cross-section, but low enough to justify the use of the small-$x$ dynamics. By inspecting the measurements in Ref.\cite{H1:2009cml}, it shows that there are only two sets of data ($Q^2=3.3~\mathrm{GeV}^2$ and $Q^2=6.6~\mathrm{GeV}^2$) simultaneously measured for the coherent and incoherent exclusive $\rho$ productions. Thus, we decide to use these two sets of data to verify the refined hot spot model.

We calculate the exclusive $\rho$ production by using the same group of parameters which are used to describe the exclusive $J/\Psi$ production data, except that the $J/\Psi$ wave function in Eq.(\ref{dipamp}) is replaced by the wave function of $\rho$. The relevant results are shown in Fig.\ref{fig6}. The left hand panels of Fig.\ref{fig6} represent the results calculated with $\mathrm{B_u}\geq\mathrm{B_d}$, while the right panels demonstrate the results computed with $\mathrm{B_u}<\mathrm{B_d}$. One can see that our refined hot spot model calculation results of the exclusive $\rho$ productions in the left hand panels are compatible with the HERA data, while the corresponding results in right hand panels are not. This outcome means that the exclusive $\rho$ production data at HERA also support the result obtained in the exclusive $J/\Psi$ production, which shows the constituent up quark having more gluons around it than the down quark at high energies. From Fig.\ref{fig6}, one can see that the model calculations do not give equally good descriptions to the HERA data as compared to $J/\Psi$ cases, especially in the incoherent process. It was found that a slightly larger constituent quark width is needed to give a steeper $|t|$ slope\cite{Mantysaari:2016jaz}. However, such a change is not favored by the exclusive $J/\Psi$ production cross-section which is good under theoretical control. As discussed above, we expect that the refined hot spot model is less reliable in the exclusive $\rho$ production process as compared the exclusive $J/\Psi$ production process due to non-perturbative contributions from large dipoles even at relatively large virtuality.

\section{conclusions and discussions}
We have calculated the coherent and incoherent exclusive vector meson production cross-sections with the refined hot spot model, in order to study the shape of the constituent quarks of the proton at high energies. The individual widths of the constituent up and down quarks are used in the event-by-event computation of the exclusive vector meson productions, instead of using the average width of the constituent quarks. We find that our refined hot spot model with the up quark width larger than or equal to the down quark one ($\mathrm{B_u}\geq\mathrm{B_d}$) can reproduce the exclusive vector meson productions at HERA, while the model calculations in other cases ($\mathrm{B_u}<\mathrm{B_d}$ or $\mathrm{B_u}\neq \mathrm{B^\prime_u}\neq\mathrm{B_d}$) cannot describe the HERA data, especially at large $|t|$. Our findings seem to indicate that the constituent up quark of the proton has more gluons surrounding it than the down quark at high energies. In order to verify that the findings are independent of model details, we use contribution weight to study the contributions of the constituent up quark and down quark to the incoherent $J/\Psi$ production cross-section. We find that the results calculated with $\mathrm{W_u}\geq\mathrm{W_d}$ can well reproduce the HERA data, while the results computed with $\mathrm{W_u}<\mathrm{W_d}$ are not consistent with the HERA data, which verify the finding just mentioned above. To further test our findings, we investigate the exclusive $\rho$ production with the same sets of parameters as used in computing exclusive $J/\Psi$ productions except the wave function replaced by the $\rho$ one. We also find that the refined hot spot model with $\mathrm{B_u}\geq\mathrm{B_d}$ is the right one favored by the HERA data of the exclusive $\rho$ production. The results calculated by the refined hot spot model with $\mathrm{B_u}<\mathrm{B_d}$ or $\mathrm{B_u}\neq \mathrm{B^\prime_u}\neq\mathrm{B_d}$ cannot reproduce the $\rho$ production data, especially at large $|t|$. Based on the above discussions, one can see that our analysis can provide an access to resolve the detailed internal structure of the proton.

In this paper, our studies are based on the exclusive vector meson production measured at HERA energy. More measurements from future Electron Ion Collider (EIC) in the USA\cite{AbdulKhalek:2021gbh}, Large Hadron electron Collider (LHeC) at CERN\cite{LHeC:2020van} and Electron-ion collider in China (EicC)\cite{Anderle:2021wcy} will give more accurate data and further dip into the small-$x$ dynamic regions, which will provide an unprecedented opportunity to access the detailed structure of the proton. Moreover, our calculations can be improved by performing the QCD evolution (dipole scattering amplitude) via the collinearly-improved Balitsky-Kovchegove equation whose Coulomb tails are strongly suppressed by the resumed kernel and can give the numerical version of the impact parameter dependent dipole scattering amplitude\cite{Cepila:2018faq}, which is our next step work.


\begin{acknowledgments}
This work is supported by the National Natural Science Foundation of China under Grant Nos.12165004, and 11947119; the Guizhou Provincial Basic Research Program (Natural Science) under Grant No.QKHJC-ZK[2023]YB027; the Education Department of Guizhou Province under Grant No.QJJ[2022]016; the National Key Research and Development Program of China under Grant Nos.2018YFE0104700, and CCNU18ZDPY04.
\end{acknowledgments}

\bibliographystyle{JHEP-2modlong}
\bibliography{refs}

\end{document}